# Vortex solitons in twisted circular waveguide arrays

Liangwei Dong[1*], Yaroslav V. Kartashov,[2,3] Lluis Torner,[2,4] and Albert Ferrando[5]

[1]*Department of Physics, Shaanxi University of Science & Technology, Xi'an 710021, China*
[2]*ICFO-Institut de Ciències Fotòniques, The Barcelona Institute of Science and Technology, 08860 Castelldefels (Barcelona), Spain*
[3]*Institute of Spectroscopy, Russian Academy of Sciences, 108840, Troitsk, Moscow, Russia*
[4]*Universitat Politècnica de Catalunya, 08034, Barcelona, Spain*
[5]*Institut de Ciència de Materials (ICMUV), Universitat de València, C/ Catedrático José Beltrán, 2, E-46980 Paterna, Spain*
*\*dongliangwei@sust.edu.cn*

We address the formation of topological states in twisted circular waveguide arrays and find that twisting leads to important differences of the fundamental properties of new vortex solitons with opposite topological charges that arise in the nonlinear regime. We find that such system features *the rare property that clockwise and counter-clockwise vortex states are nonequivalent*. Focusing on arrays with $\mathcal{C}_{6v}$ discrete rotation symmetry, we find that a longitudinal twist stabilizes the vortex solitons with the lowest topological charges $m=\pm 1$, which are always unstable in untwisted arrays with the same symmetry. Twisting also leads to the appearance of instability domains for otherwise stable solitons with $m=\pm 2$ and generates vortex modes with topological charges $m=\pm 3$ that are forbidden in untwisted arrays. By and large, we establish a rigorous relation between the discrete rotation symmetry of the array, its twist direction, and the possible soliton topological charges.

PhySH Subject Headings: Solitons; Waveguide arrays

Propagation of light is strongly affected by the presence of transverse refractive index modulations [1-5], a fundamental effect that generates a variety of physical phenomena. In nonlinear media, such refractive index landscapes (in particular, periodic geometries) may support rich families of lattice solitons, which have been observed in multi-dimensional geometries [6-12]. Lattices may suppress azimuthal modulation instabilities, thus stabilizing vortex states (see reviews [13-15], experimental realizations [16-20] and theoretical proposals [9, 21-25], including in dissipative systems [26]). Vortices can also exist in structures exhibiting discrete angular rotational symmetry rather than transverse periodicity, such as modulated Bessel lattices [27-31] or photonic crystals [32]. Such structures revealed the surprising fact that the lattice symmetry (namely, the order of its discrete rotational symmetry) imposes rigorous restrictions on the topological charges of symmetric vortex solitons and determines their stability properties [32-34]. An angular pseudo-momentum theory for modes in such structures has been developed in [32-37].

Circular waveguide arrays with discrete angular rotational symmetry are important in connection with transmission and compression of the high-energy pulses and beam combining [38, 39]. They support various vortex states with similar restrictions on topological charge [40-43], gyrating [44] and stationary gap [45] and multipole [46] states on a ring, and they allow the observation of unconventional modulation instability regimes [47], and angular soliton switching [30, 48, 49]. Crucially, all such systems are characterized by full clockwise and anti-clockwise symmetry, and thus by identical properties of states with opposite topological charges. Equivalence of vortex states opposite charges was also observed in Bose-Einstein condensates [50-53] and polariton microcavities [54, 55].

A fundamentally different situation arises when the underlying structure with discrete rotational symmetry is twisted in the direction of propagation. In such systems, the helicity dramatically affects light transfer between its cores, enabling unusual diffraction management [56, 57] and topological suppression of tunneling between cores [58, 59], recently observed in linear [60] and nonlinear [61] systems. Also, in PT-symmetric ring arrays twisting allows controlling the system symmetry-breaking threshold [62-64]. Time-modulated circular arrays were used at radio frequencies to generate new vortex states [65].

Nevertheless, the impact of twisting on the properties and stability of *self-sustained* vortex states in the *nonlinear regime* has not been addressed, neither in ring-like arrays nor in rotating lattices, such as those described in [66-71]. Such rotating structures can be created using interfering nondiffracting beams [72, 73] propagating in photo-sensitive materials [74-76], or as photonic crystal fibers twisted during drawing process [77-79]. The non-degeneracy of linear vortex modes with opposite charges occurs in linear twisted photonic crystal fibers [80] and is suggested by the excitations in optically-induced structures [76], but it has not been addressed directly in the nonlinear regime.

In this Letter we will show that vortex solitons with opposite topological charges in twisted circular waveguide arrays feature different domains of existence and stability properties. We find that twisting can stabilize states that are always unstable in static structures, and that it can generate vortex states with topological charges that are forbidden in untwisted systems. We cast our findings on general discrete-symmetry grounds, so that many of our results also hold in other non-reciprocal systems with finite rotational order.

We address the paraxial propagation of light in a twisted waveguide array with focusing nonlinearity described by the dimensionless nonlinear Schrödinger equation for the field amplitude $\psi$ (for details of its derivation from Maxwell equations see [81]):

$$i\frac{\partial \psi}{\partial z} = -\frac{1}{2}\left(\frac{\partial^2 \psi}{\partial x^2} + \frac{\partial^2 \psi}{\partial y^2}\right) - |\psi|^2 \psi - \mathcal{R}(x,y,z)\psi. \qquad (1)$$

Here we normalize the $x,y$ coordinates to the characteristic transverse scale $r_0 = 10~\mu\mathrm{m}$, the propagation distance $z$ to the diffraction length $kr_0^2 \approx 1.44~\mathrm{mm}$ (assuming $\lambda=800~\mathrm{nm}$ wavelength), $k=2\pi n/\lambda$, $n\approx 1.45$ is the unperturbed refractive index of the material. The function $\mathcal{R} = p\sum_{k=1,N} e^{-[(x-x_k)^2+(y-y_k)^2]/a^2}$ describes $N$ single-mode Gaussian guides with width $a=0.5$ (5 $\mu\mathrm{m}$) and depth $p=k^2 r_0^2 \delta n / n$ defined by real refractive index contrast $\delta n$ that are placed on a ring of radius $\rho = 0.3N$. We use $p=8$ that corresponds to $\delta n \approx 9\times 10^{-4}$. The structure rotates with frequency $\omega$ in the direction of light propagation, the coordinates of the waveguides on a ring are $x_k = \rho\cos(\phi_k - \omega z)$, $y_k = \rho\sin(\phi_k - \omega z)$, where $\phi_k = 2\pi(k-1)/N$ and $\omega > 0$ corresponds to the clockwise twist. For $\omega = 0.1$ the period of rotation is $\sim 90~\mathrm{mm}$. Such twisted arrays can be inscribed with fs laser in fused silica [82,83]; a twisted version of photonic crystal fibers [11,12] with low $\delta n$ can be used too. In fused silica samples ($n_2 = 2.2 \times 10^{-20}$ m$^2$/W) dimensionless intensity $|\psi|^2 = 1$ corresponds to peak intensity $I = n|\psi|^2 / k^2 r_0^2 n_2 \approx 5 \times 10^{15}$ W/m$^2$. Under the above conditions, two-photon absorption in Eq. (1) can be neglected [82]. When $\omega = 0$, the rotational symmetry group of a lattice of order $N$ is given by the discrete point-symmetry group $\mathcal{C}_{Nv}$ corresponding to discrete rotations by the angle $\varepsilon_N = 2\pi/N$ and to specular reflections with respect to a number of planes containing the rotation axis. This symmetry dictates that in both linear and nonlinear regimes the system supports only vortex states with topological charges $0 < |m| < N/2$ (for even $N$) and that the properties of $+m$ and $-m$ states are identical [33]. This picture changes dramatically in rotating arrays. To show this, we cast Eq. (1) in the rotating coordinate frame $x' = x\cos(\omega z) + y\sin(\omega z)$, $y' = y\cos(\omega z) - x\sin(\omega z)$, where the array profile $\mathcal{R}$ is independent of the propagation distance $z$:

$$i\frac{\partial \psi}{\partial z} = -\frac{1}{2}\left(\frac{\partial^2 \psi}{\partial x^2} + \frac{\partial^2 \psi}{\partial y^2}\right) + \omega \mathcal{L}_z \psi - |\psi|^2 \psi - \mathcal{R}(x,y)\psi, \quad (2)$$

and rotating operator $\omega \mathcal{L}_z = i\omega(x\partial/\partial y - y\partial/\partial x)$ is introduced. Here primes in coordinates are omitted for simplicity. The rotating operator term $\omega \mathcal{L}_z = -i\omega \partial/\partial \theta$ does not change the original rotational symmetry since the derivative is also invariant under any discrete rotation $\theta \to \theta + 2\pi/N$. Thus, the rotating linear Hamiltonian $\mathcal{H}(\omega) = \mathcal{H}(0) + \omega \mathcal{L}_z = -(1/2)\nabla^2 - \mathcal{R} + \omega \mathcal{L}_z$ is also invariant under discrete rotations of the point group $\mathcal{C}_N$. The rotating operator $\omega \mathcal{L}_z$ is self-adjoint, which ensures that the eigenvalues $b(\omega)$ of the rotating Hamiltonian $\mathcal{H}(\omega) = \mathcal{H}(0) + \omega \mathcal{L}_z$ are also real. However, under complex conjugation the rotating operator term changes sign: $\omega \mathcal{L}_z \xrightarrow{C} \omega \mathcal{L}_z^* = -\omega \mathcal{L}_z$. Thus, the Hamiltonian operator $\mathcal{H}(\omega)$ is self-adjoint but not invariant under conjugation. In fact, complex conjugation links the rotating Hamiltonian at frequency $\omega$ with its counterpart at frequency $-\omega$: $\mathcal{H}^*(\omega) = \mathcal{H}(-\omega)$. Under a "time reversal" transformation $\mathcal{T}$ ($z \to -z$), the sign of rotation changes, so that $\omega \to -\omega$. Thus, $\mathcal{T}$ produces the same effect as complex conjugation, since $\mathcal{T}^{-1}\mathcal{H}(\omega)\mathcal{T} = \mathcal{H}(-\omega) = \mathcal{H}^*(\omega)$.

Due to the $\mathcal{C}_N$ invariance of $\mathcal{H}(\omega)$ for all $\omega$, the eigenfunctions of the rotating Hamiltonian are angular Bloch modes with well-defined orbital angular pseudo-momentum (OAPM) $m$ [35]. The OAPM $m$ labels every Hamiltonian eigenfunction. Besides, $m$ sets the on-axis topological charge of the angular mode [36]. Angular modes can be written in the form $\psi_{m,\omega}(r,\theta,z) = u_{m,\omega}(r,\theta)e^{im\theta + ib_m z}$, where $u_{m,\omega}$ is the angular Bloch function, which is periodic in angle $\theta$: $u_{m,\omega}(r, \theta + 2\pi/N) = u_{m,\omega}(r,\theta)$. As in standard Bloch theory, for every set of modes defined by their OAPM $m$ we can write their corresponding eigenvalue equation $\mathcal{H}_m(\omega) u_{m,\omega} = -b_m(\omega) u_{m,\omega}$ for the angular Bloch function $u_{m,\omega}$ of a stationary state. The symmetries of the reduced Hamiltonian $\mathcal{H}_m(\omega)$ obtained from the $\mathcal{H}(\omega)$ after the substitution of $\psi_{m,\omega}$ determine the properties of the angular Bloch mode functions and their propagation constants $b_m(\omega)$. The reduced Hamiltonian is also self-adjoint, which guarantees that all $b_m(\omega)$ are real. However, as the original Hamiltonian, $\mathcal{H}_m(\omega)$ is not real and also gets transformed by time reversal as $\mathcal{T}^{-1}\mathcal{H}_m(\omega)\mathcal{T} = \mathcal{H}_{-m}(-\omega) = \mathcal{H}_m^*(\omega)$. As a consequence, angular Bloch functions and propagation constants of the rotating Hamiltonian must fulfill:

$$u_{m,\omega}^* = u_{-m,-\omega}, \quad b_m(\omega) = b_{-m}(-\omega). \quad (3)$$

Thus, the symmetry pattern of the system changes when we pass from $\omega = 0$ to $\omega \neq 0$. This change in the symmetry of the Hamiltonian explains the splitting of the OAPM doublets (the modes with identical $b$ corresponding to opposite values of $m \neq 0$) existing in the $\omega = 0$ case once we turn on the rotating term. By introducing this term, we break $\mathcal{T}$ symmetry and the pair of modes $u_{m,\omega}$ and $u_{-m,\omega}$ stops being degenerate since this degeneracy only occurs when the full Hamiltonian $\mathcal{H}(\omega)$ is real, i.e., when $\omega = 0$. This result is an explicit demonstration of a general property in group theory. The representations of the point group $\mathcal{C}_N$ are one-dimensional, which means that they are *not* degenerate in general. However, for $\omega = 0$ the $m$ and $-m$ representations are degenerate and complex conjugated of each other, as shown in Fig. 1(a). This is so because of the invariance under $\mathcal{T}$ of the non-rotating Hamiltonian: $\mathcal{H}^*(0) = \mathcal{H}(0)$ [37]. The presence of the rotating term breaks time-reversal symmetry, and therefore produces the splitting of the $m$ and $-m$ doublets with the same value of $\omega$, which are no longer degenerate. However, this splitting is not arbitrary since a new degeneracy appears. The $(m,\omega)$ and $(-m,-\omega)$ modes are now degenerate ones since they are connected by $\mathcal{T}$, as seen in Eq. (3) and in Fig. 1(b). In this sense, $\omega$ plays a role analogous to a magnetic field along the $z$-direction since time reversal $\mathcal{T}$ induces the $\omega \to -\omega$ transformation, as in the magnetic case.

Breaking time reversal symmetry is connected to non-reciprocity, as elaborated in [84]. Non-reciprocity is achieved by breaking time-reversal symmetry with an external bias field $\vec{F}_0$, which has to be odd under time-reversal. In our case the bias field is the twist conferred by the rotation frequency $\vec{\omega} = \omega\hat{k}$ to the waveguide array in such a way the twisting angle along the $z$-direction is given by $\vec{\theta}(z) = \omega z \hat{k}$ (see the first section of [85] for details). Nonreciprocity manifest in the inequivalent dynamics (different Hamiltonian) that evolving states experience for forward $\mathcal{H}(\omega)$ and backward $\mathcal{H}(-\omega)$ propagation. This is a general feature of nonreciprocal lossless systems under the action of any bias field $\vec{F}_0$ odd under time-reversal. Therefore, any linear or nonlinear lossless scalar system owning discrete rotational symmetry in the presence of a bias field $\vec{F}_0$ breaking time-reversal/mirror-reflection symmetry should present the same qualitative spectral properties.

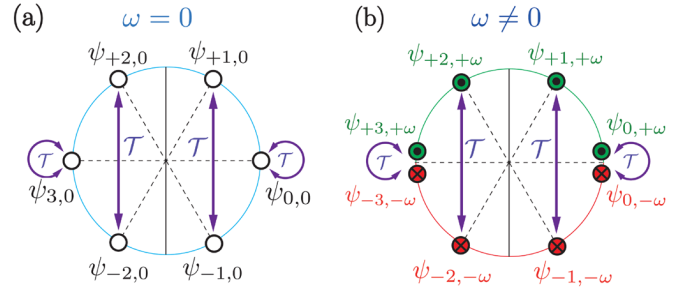

Fig. 1. Angular Bloch modes $\psi_m$ as representations of the point group $\mathcal{C}_6$ for exemplary array with $N = 6$. They are described as the roots of unity $e^{im\pi/3}$ in terms of their OAPM $m$. (a) For $\omega = 0$ (non-rotating case) the $m = \pm 1, \pm 2$ modes form degenerate doublets due to time reversal symmetry. (b) For $\omega \neq 0$, time reversal is broken, which now connects an $m$ mode with its partner $-m$ with opposite value of $\omega$. A frequency value $-\omega$ is represented by a red cross whereas a frequency value $\omega$ is indicated by a green dot.

The modes with $m = 0$ and $m = 3$ deserve a particular consideration. Group theory tells that these modes (recall that $m = +3$ and $m = -3$ corresponds to the same state) are singlets since they belong to one-dimensional representations of $\mathcal{C}_6$. When $\omega = 0$ these modes are invariant, up to a sign, under complex conjugation or, equivalently, under $\mathcal{T}$. Accordingly, they are real functions of a multipole type. When $\omega \neq 0$, they are still singlets, but become complex functions verifying $\psi_{0,\omega}^* = \psi_{0,-\omega}$ and $\psi_{3,\omega}^* = \psi_{-3,-\omega}$. In the case of $m=3$ the rotation turns the real multipole singlet of the non-rotating case $\psi_{3,0}$ into a single complex mode that shows different behavior depending on the sign of $\omega$: $\psi_{3,\omega} = e^{+i3\theta} u_{3,|\omega|}$ if $\omega > 0$ and $\psi_{-3,\omega} = e^{-i3\theta} u_{3,|\omega|}^*$ if $\omega < 0$. Since the value of the OAPM $m$ defines the on-axis topological charge, the singlet $|m|=3$ appears as an on-axis vortex of charge $+3$ for positive $\omega$ and as a vortex of charge $-3$ for negative $\omega$ - thus rotation generates vortex modes that are forbidden at $\omega = 0$.

Using the above symmetry properties one can predict a general functional form of the $b_m(\omega)$ dependence for linear modes. Due to the $\mathcal{C}_N$ symmetry of the rotating array the propagation constant has to fulfill also the following periodicity property:

$$b_{m+N}(\omega) = b_m(\omega) \qquad (4)$$

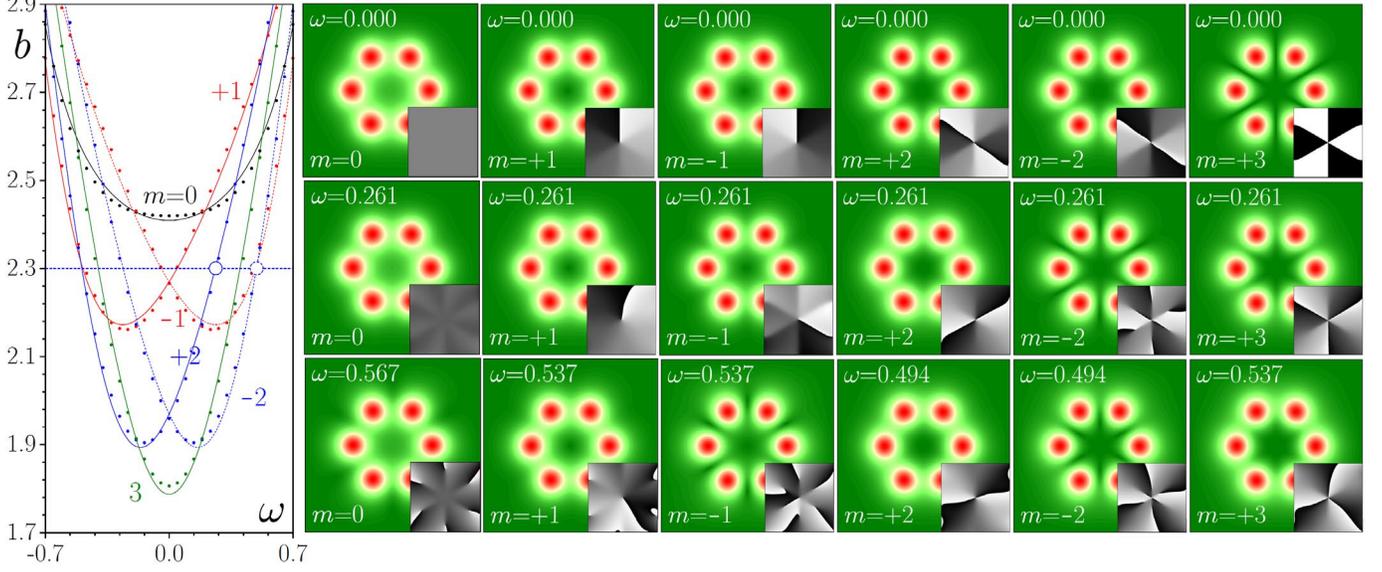

Fig. 2. Exact numerically calculated eigenvalues (lines) and their analytical approximation (dots) for linear modes of the rotating structure with $N=6$ waveguides vs $\omega$ and examples of modes ($|\psi|$ distributions and $\arg(\psi)$ distributions in the insets) with different $m,\omega$ values. Open blue circles indicate crossings of the horizontal dashed line $b=2.3$ with $m=\pm 2$ mode families that occur at different frequencies.

An example of the *ansatz* compatible with symmetries (3) and (4) is $b_m(\omega) = \varepsilon(\omega) + c(\omega)\cos(2\pi m/N) + \omega s(\omega)\sin(2\pi m/N)$, where $\varepsilon(\omega) = \sum_r \varepsilon_r \omega^{2r}$, $c(\omega) = \sum_r c_r \omega^{2r}$, and $s(\omega) = \sum_r s_r \omega^{2r}$ are even polynomial functions of $\omega$. In Fig. 2 (left) we compare for a representative case of $N=6$ the exact numerically calculated dependencies $b_m(\omega)$ (lines) with the above *ansatz* [dots, obtained by adjusting the coefficients $\varepsilon_r, c_r, s_r$ in the *ansatz*, where we kept terms up to $\mathcal{O}(\omega^4)$]. The agreement is remarkably good for all modes. Note that our *ansatz* is valid also for the nonlinear case since all symmetry arguments hold for the nonlinear equation as well. Figure 2 also illustrates transformation of the field modulus and phase distributions in exact linear modes with increase of $\omega$. For modes with $m<0$ one observes the appearance of several off-center phase singularities that gradually approach the center of the array with increase of $\omega$. Variation of $\omega$ substantially changes angular modulation depth in field modulus distributions. Transformation of multipole states into vortex-carrying ones at $\omega \neq 0$ is illustrated too (right column).

Next we consider *vortex solitons* by solving Eq. (2) with cubic nonlinearity included. Solitons are sought in the form $\psi = q(x,y)e^{ibz}$. Their properties are summarized in Fig. 3 and 4. We first fix propagation constant $b$ and increase rotation frequency $\omega$. Soliton power $U = \iint |\psi|^2 dx dy$ decreases with $\omega$ for $m=+1,+2,\pm 3$ and varies nonmonotonically for $m=-1,-2$ [Fig. 3(a),(c),(e)]. In all cases, solitons transform into linear modes at critical frequencies $\omega_{\mathrm{cr}}$ different for positive and negative topological charges. Critical frequencies can be determined from the linear spectrum in Fig. 2 from the intersections of the linear dispersion curves $b_{\pm m}(\omega)$ with horizontal line corresponding to selected soliton propagation constant (see horizontal dashed blue line, for example). Due to asymmetry of linear dependencies $b_m(\omega)$ for $m<0$ the point of intersection is located at larger frequency than for $m>0$, hence solitons with negative charges cease to exist at larger frequency values at $\omega > 0$ side. The larger is the soliton propagation constant, the larger is the interval of rotation frequencies, where it can exist [Fig. 3(e)]. However, when rotation frequency becomes too large, the waveguides become leaky (for our parameters this occurs for $|\omega| > 1$) and it is necessary to further increase array depth $p$ to obtain steadily rotating states.

To test stability of vortex solitons we substitute their perturbed profiles $\psi = (q + u^{\lambda z} + v^* e^{\lambda^* z})e^{ibz}$ with $u,v \ll q$ into Eq. (2), linearize it and solve corresponding linear eigenvalue problem to obtain perturbation growth rates $\lambda = \lambda_{\mathrm{re}} + i\lambda_{\mathrm{im}}$, whose dependencies on $\omega$ are presented in Fig. 3(b),(d),(f). Solitons are stable when $\lambda_{\mathrm{re}} = 0$. Remarkably, for $m=\pm 1$ maximal perturbation growth rate vanishes above certain critical rotation frequency, meaning that twist stabilizes vortex solitons with the lowest topological charges that are usually unstable in static systems with $\mathcal{C}_6$ discrete rotation symmetry [34, 20]. At the same time, twist may also destabilize some of the solitons that were stable at $\omega = 0$. This is seen for $m=+2$ and $m=+3$ states that feature instability islands (they are shown by the red color, while all stable branches in Fig. 3 are shown black). Notice that for both $m=\pm 1$ and $m=\pm 2$ solitons stability properties are different due to structure twist. In contrast, solitons with $m=\pm 3$ always feature the same stability intervals, and $m=0$ solitons are always stable. Increasing propagation constant at fixed $\omega$ leads to growth of soliton power, see Fig. 4(a) (the curves for different $m$ values are similar, so we show them only for $m=\pm2$). Increasing $b$ usually leads to stabilization of solitons with the highest topological charges and destabilization of states with the lowest charges [Fig. 4(b)].

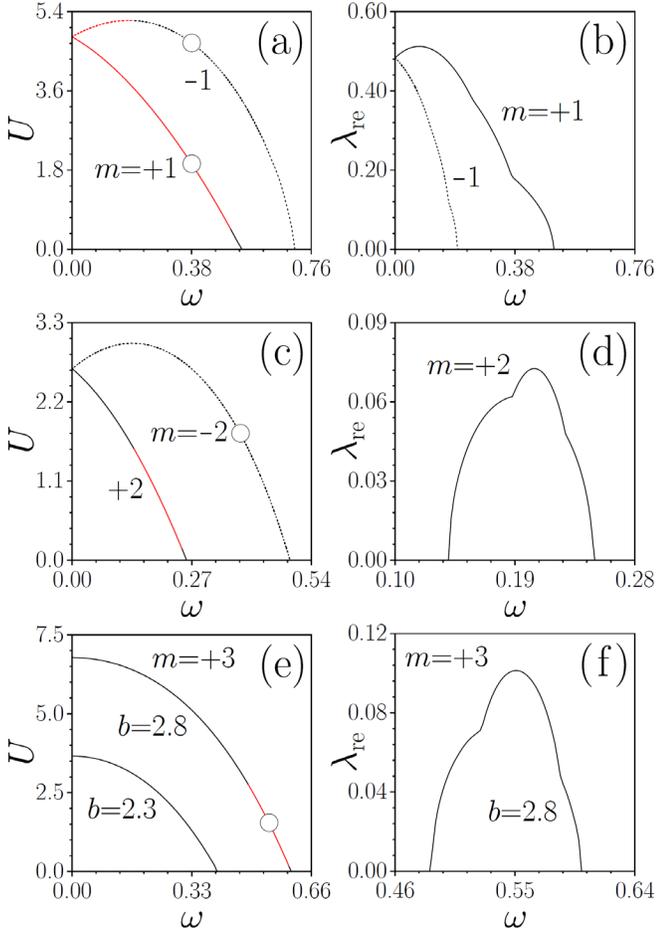

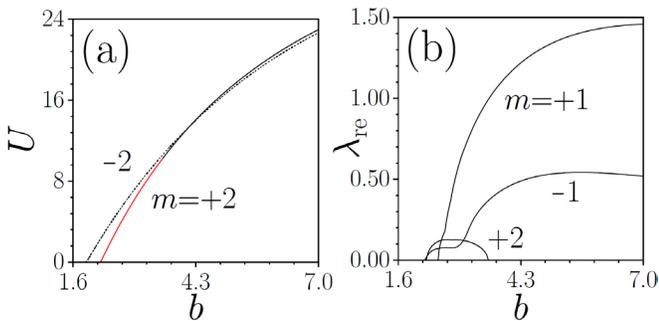

Fig. 3. Power (left) and maximal perturbation growth rate (right) versus rotation frequency $\omega$ for $m=\pm1$ solitons at $b=2.8$ (a),(b), $m=\pm2$ solitons at $b=2.3$ (c),(d), and $m=3$ solitons at $b=2.3$ and $b=2.8$ (e),(f). Stable branches are shown black, unstable ones are shown red. Dots correspond to solitons, whose propagation dynamics is depicted in Fig. 5.

Fig. 4. (a) Power vs propagation constant $b$ for $m=\pm2$ solitons at $\omega=0.2$. (b) Maximal perturbation growth rate vs $b$ for all solitons with $m=\pm1,\pm2,+3$ at $\omega=0.2$.

Figure 5 shows examples of stable propagation of vortex solitons with topological charges $m=-1$ [Fig. 5(a)] and $m=-2$ [Fig. 5(c)] corresponding to the dots in Fig. 3 in the presence of broadband noise added into the input field distributions. Simulations were performed in the nonrotating coordinate frame, using Eq. (1). These structures show stable persistent rotation over huge distances. When vortex solitons are unstable, instability is manifested in development of the azimuthal modulation and irregular field oscillations in all waveguides [Fig. 5(b) and 5(d)]. The generality of our findings is supported by the analysis of twisted arrays with other types of discrete rotational symmetry, such as $\mathcal{C}_8$ and $\mathcal{C}_{10}$ (see [85]), where one also observes rotation-induced splitting of linear OAPM doublets existing at $\omega=0$ and formation of vortex modes that are forbidden at $\omega=0$, as well as rotation-induced stabilization of vortex solitons with lowest topological charges and destabilization of higher-charge states.

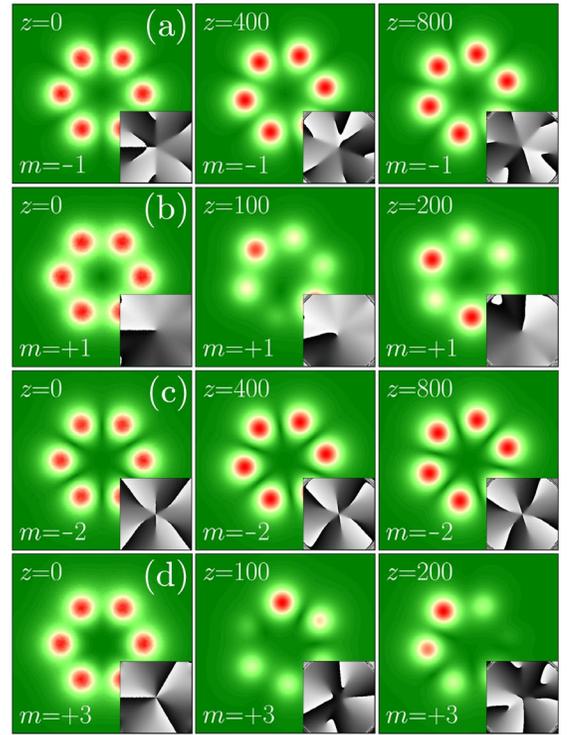

Fig. 5. Propagation of stable vortex solitons with $m=-1$, $b=2.8$, $\omega=0.38$ (a) and $m=-2$, $b=2.3$, $\omega=0.38$ (c), and decay of the unstable states with $m=+1$, $b=2.8$, $\omega=0.38$ (b), and $m=+3$, $b=2.8$, $\omega=0.54$ (d).

In summary, we have shown that twisting waveguide arrays with discrete rotational symmetry profoundly affects the domains of existence and stability properties of nonlinear vortices. We found that twisting breaks the equivalence between states with equal but opposite topological charges. Our results are based on general group-theory arguments, which hold for a wide variety of physical systems. The underlying optical system is readily realizable experimentally and enriches the class of settings that feature nonequivalent clockwise and counter-clockwise vortical currents [86-88]. In particular, dynamically varying $\mathcal{C}_{Nv}$ potentials can be created in clouds of cold atoms, Bose-Einstein condensates, optical waveguide arrays, photonic crystal fibers, atomic vapors, and polariton condensates.

Y.V.K., A.F. and L.T. acknowledge grant CEX2019-000910-S funded by MCIN/AEI/ 10.13039/501100011033, Fundació Cellex, Fundació Mir-Puig, and Generalitat de Catalunya (CERCA, AGAUR). A.F. thanks the support of MCIN of Spain through the project PID2020-120484RB-I00 and Generalitat Valenciana, Spain (grant PROMETEO/2021/082).

## REFERENCES

1. F. Lederer, G. I. Stegeman, D. N. Christodoulides, G. Assanto, M. Segev, Y. Silberberg, "Discrete solitons in optics," Phys. Rep. 463, 1 (2008).
2. Y. V. Kartashov, V. A. Vysloukh, and L. Torner, "Soliton shape and mobility control in optical lattices," Prog. Opt. 52, 63 (2009).
3. Y. V. Kartashov, G. Astrakharchik, B. Malomed, and L. Torner, "Frontiers in multidimensional self-trapping of nonlinear fields and matter," Nat. Rev. Phys. 1, 185 (2019).
4. D. Mihalache, "Localized structures in optical and matter-wave media: A selection of recent studies," Rom. Rep. Phys. 73, 403 (2021).
5. M. Barbuto, A. Alu, F. Bilotti, A. Toscano, "Dual-circularly polarized topological patch antenna with pattern diversity," IEEE Access **9**, 48769 (2021).
6. N. K. Efremidis, S. Sears, D. N. Christodoulides, J. W. Fleischer, M. Segev, "Discrete solitons in photorefractive optically induced photonic lattices," Phys. Rev. E 66, 046602 (2002).
7. J. W. Fleischer, M. Segev, N. K. Efremidis, D. N. Christodoulides, "Observation of two-dimensional discrete solitons in optically induced nonlinear photonic lattices," Nature 422, 147 (2003).
8. D. Neshev, E. Ostrovskaya, Y. Kivshar, W. Krolikowski, "Spatial solitons in optically induced gratings," Opt. Lett. **28**, 710 (2003).
9. J. Yang, Z. H. Musslimani, "Fundamental and vortex solitons in a two-dimensional optical lattice," Opt. Lett. **28**, 2094 (2003).
10. N. K. Efremidis, J. Hudock, D. N. Christodoulides, J. W. Fleischer, O, Cohen, M. Segev, "Two-dimensional optical lattice solitons," Phys. Rev. Lett. **91**, 213906 (2003).
11. S. Minardi, F. Eilenberger, Y. V. Kartashov, A. Szameit, U. Röpke, J. Kobelke, K. Schuster, H. Bartelt, S. Nolte, L. Torner, F. Lederer, A. Tünnermann, and T. Pertsch, "Three-dimensional light bullets in arrays of waveguides," Phys. Rev. Lett. 105, 263901 (2010).
12. F. Eilenberger, K. Prater, S. Minardi, R. Geiss, U. Röpke, J. Kobelke, K. Schuster, H. Bartelt, S. Nolte, A. Tünnermann, and T. Pertsch, "Observation of discrete, vortex light bullets," Phys. Rev. X 3, 041031 (2013).
13. A. S. Desyatnikov, Y. S. Kivshar, and L. Torner, "Optical vortices and vortex solitons," Prog. Opt. 47, 291 (2005).
14. B. A. Malomed, "Vortex solitons: Old results and new perspectives," Physica D 399, 108 (2019).
15. A. Pryamikov, L. Hadzievski, M. Fedoruk, S. Turitsyn and A. Aceves, "Optical vortices in waveguides with discrete and continuous rotational symmetry," Journal of the European Optical Society-Rapid Publications 17, 23 (2021).
16. D. N. Neshev, T. J. Alexander, E. A. Ostrovskaya, Y. S. Kivshar, H. Martin, I. Makasyuk, and Z. Chen, "Observation of discrete vortex solitons in optically induced photonic lattices," Phys. Rev. Lett. 92, 123903 (2004).
17. J. W. Fleischer, G. Bartal, O. Cohen, O. Manela, M. Segev, J. Hudock, and D. N. Christodoulides, "Observation of vortex-ring discrete solitons in 2D photonic lattices," Phys. Rev. Lett. 92, 123904 (2004).
18. G. Bartal, O. Manela, O. Cohen, J. W. Fleischer, and M. Segev, "Observation of second-band vortex solitons in 2d photonic lattices," Phys. Rev. Lett. 95, 053904 (2005).
19. B. Terhalle, T. Richter, A. S. Desyatnikov, D. N. Neshev, W. Krolikowski, F. Kaiser, C. Denz, and Y. S. Kivshar, "Observation of multivortex solitons in photonic lattices," Phys. Rev. Lett. 101, 013903 (2008).
20. B. Terhalle, T. Richter, K. J. H. Law, D. Gories, P. Rose, T. J. Alexander, P. G. Kevrekidis, A. S. Desyatnikov, W. Krolikowski, F. Kaiser, C. Denz, and Y. S. Kivshar, "Observation of double-charge discrete vortex solitons in hexagonal photonic lattices," Phys. Rev. A 79, 043821 (2009).
21. B. A. Malomed and P. G. Kevrekidis, "Discrete vortex solitons," Phys. Rev. E 64, 026601 (2001).
22. K. J. H. Law, P. G. Kevrekidis, T. J. Alexander, W. Krolikowski, Y. S. Kivshar, "Stable higher-charge discrete vortices in hexagonal optical lattices," Phys. Rev. A 79, 025801 (2009).
23. P. G. Kevrekidis, B. A. Malomed, Z. Chen, and D. J. Frantzeskakis, "Stable higher-order vortices and quasivortices in the discrete nonlinear Schrödinger equation," Phys. Rev. E 70, 056612 (2004).
24. T. J. Alexander, A. A. Sukhorukov, Y. S. Kivshar, "Asymmetric vortex solitons in nonlinear periodic lattices," Phys. Rev. Lett. 93, 063901 (2004).
25. M. Öster and M. Johansson, "Stable stationary and quasi-periodic discrete vortex breathers with topological charge S=2," Phys. Rev. E 73, 066608 (2006).
26. M. Honari-Latifpour, J. Ding, M. Barbuto, S. Takei, and M.-A. Miri, "Self-organized vortex and antivortex patterns in laser arrays," Phys. Rev. Applied **16**, 054010 (2021).
27. Y. V. Kartashov, V. A. Vysloukh, and L. Torner, "Rotary solitons in Bessel optical lattices," Phys. Rev. Lett. 93, 093904 (2004).
28. Y. V. Kartashov, V. A. Vysloukh, L. Torner, "Stable ring-profile vortex solitons in Bessel optical lattices," Phys. Rev. Lett. 94, 043902 (2005).
29. X. S. Wang, Z. G. Chen, P. G. Kevrekidis, "Observation of discrete solitons and soliton rotation in optically induced periodic ring lattices," Phys. Rev. Lett. 96, 083904 (2006).
30. Y. V. Kartashov, A. A. Egorov, V. A. Vysloukh, and L. Torner, "Stable soliton complexes and azimuthal switching in modulated Bessel optical lattices," Phys. Rev. E 70, 065602(R) (2004).
31. R. Fischer, D. N. Neshev, S. Lopez-Aguayo, A. S. Desyatnikov, A. A. Sukhorukov, W. Krolikowski, Yu. S. Kivshar, "Observation of light localization in modulated Bessel optical lattices," Opt. Express 14, 2825 (2006).
32. A. Ferrando, M. Zacares, M. A. Garcia-March, J. A. Monsoriu, and P. F. de Cordoba, "Vortex transmutation," Phys. Rev. Lett. **95**, 123901 (2005).
33. A. Ferrando, M. Zacares, and M. A. Garcia-March, "Vorticity cutoff in nonlinear photonic crystals," Phys. Rev. Lett. **95**, 043901 (2005).
34. Y. V. Kartashov, A. Ferrando, A. A. Egorov, and L. Torner, "Soliton topology versus discrete symmetry in optical lattices," Phys. Rev. Lett. **95**, 123902 (2005).
35. A. Ferrando, "Discrete-symmetry vortices as angular Bloch modes," Phys. Rev. E **72**, 036612 (2005).
36. M. A. García-March, A. Ferrando, M. Zacarés, S. Sahu, D. E. Ceballos Herrera, "Symmetry, winding number, and topological charge of vortex solitons in discrete-symmetry media," Phys. Rev. A **79**, 053820 (2009).
37. M. Hamermesh, "Group theory and its application to physical problems" (First ed.). Reading, Massachusetts: Addison-Wesley (1964).
38. A. M. Rubenchik, I. S. Chekhovskoy, M. P. Fedoruk, O. V. Shtyrina, and S. K. Turitsyn, "Nonlinear pulse combining and pulse compression in multi-core fibers," Opt. Lett. **40**, 721 (2015).
39. C. N. Alexeyev, A. V. Volyar, M. A. Yavorsky, "Linear azimuthons in circular fiber arrays and optical angular momentum of discrete optical vortices," Phys. Rev. A. 80, 063821 (2009).
40. D. Leykam, B. Malomed, and A. S. Desyatnikov, "Composite vortices in nonlinear circular waveguide arrays," J Opt 15, 044016 (2013).
41. L. Hadzievski, A. Maluckov, A. Rubenchik, and S. Turitsyn, "Stable optical vortices in nonlinear multicore fibers," Light Sci. Appl. 4, 314 (2015).
42. A. Radosavljevic, A. Danicic, J. Petrovic, A. Maluckov, L. Hadzievski, "Coherent light propagation through multicore optical fibers with linearly coupled cores," J. Opt. Soc. Am. B 32, 2520 (2015).
43. L. Dong, H. J. Li, C. M. Huang, S. S. Zhong, C. Y. Li, "Higher-charged vortices in mixed linear-nonlinear circular arrays," Phys. Rev. A 84, 043830 (2011).
44. I. V. Barashenkov and D. Feinstein, "Gyrating solitons in a necklace of optical waveguides," Phys. Rev. A 103, 023532 (2021).
45. Y. V. Kartashov, B. A. Malomed, V. A. Vysloukh, and L. Torner, "Gap solitons on a ring," Opt. Lett. 33, 2949 (2008).
46. Y. V. Kartashov, B. A. Malomed, V. A. Vysloukh, and L. Torner, "Stabilization of multibeam necklace solitons in circular arrays with spatially modulated nonlinearity," Phys. Rev. A 80, 053816 (2009).
47. C. Maitland, D. Faccio, and F. Biancalana, "Modulation instability of discrete angular momentum in coupled fiber rings," J. Opt. 21, 065504 (2019).
48. W. Krolikowski, U. Trutschel, M. Cronin-Golomb, and C. Schmidt-Hattenberger, "Solitonlike optical switching in a circular fiber array," Opt. Lett. 19, 320 (1994).

49. A. S. Desyatnikov, M. R. Dennis, and A. Ferrando, "All-optical discrete vortex switch," Phys. Rev. A 83, 063822 (2011).
50. J. Denschlag, J. E. Simsarian, D. L. Feder, Charles W. Clark, L. A. Collins, J. Cubizolles, L. Deng, E. W. Hagley, K. Helmer-son, W. P. Reinhardt, S. L. Rolston, B. I. Schneider, and W. D. Phillips, "Generating solitons by phase engineering of a Bose-Einstein condensate," Science 287, 97 (2000).
51. A. E. Leanhardt, A. Gorlitz, A. P. Chikkatur, D. Kielpinski, Y. Shin, D. E. Pritchard, and W. Ketterle, "Imprinting vortices in a Bose-Einstein condensate using topological phases," Phys. Rev. Lett. 89, 190403 (2002).
52. J. E. Williams and M. J. Holland, "Preparing topological states of a Bose–Einstein condensate," Nature 401, 568 (1999).
53. A. L. Fetter, "Rotating trapped Bose-Einstein condensates," Rev. Mod. Phys. 81, 647 (2009).
54. K. G. Lagoudakis, M. Wouters, M. Richard, A. Baas, I. Carusotto, R. Andre, L. E. S. I. Dang, B. Deveaud-Pledran, "Quantized vortices in an exciton-polariton condensate," Nat. Phys. 4, 706 (2008).
55. D. Sanvitto, F. M. Marchetti, M. H. Szymanska, G. Tosi, M. Baudisch, F. P. Laussy, D. N. Krizhanovskii, M. S. Skolnick, L. Marrucci, A. Lemaitre, J. Bloch, C. Tejedor, L. Vina, "Persistent currents and quantized vortices in a polariton superfluid," Nat. Phys. 6, 527 (2010).
56. S. Longhi, "Bloch dynamics of light waves in helical optical waveguide arrays," Phys. Rev. B 76, 195119 (2007).
57. S. Longhi, "Light transfer control and diffraction management in circular fibre waveguide arrays," Journal of Physics B: Atomic, Molecular and Optical Physics 40, 4477 (2007).
58. M. Ornigotti, G. Della Valle, D. Gatti, and S. Longhi, "Topological suppression of optical tunneling in a twisted annular fiber," Phys. Rev. A 76, 023833 (2007).
59. C. Castro-Castro, Y. Shen, G. Srinivasan, A. B. Aceves, P. G. Kevrekidis, "Light dynamics in nonlinear trimers and twisted multicore fibers," J. Nonlinear Opt. Phys. Mater. 25, 1650042 (2016).
60. M. Parto, H. Lopez-Aviles, J. E. Antonio-Lopez, M. Khajavikhan, R. Amezcua-Correa, D. N. Christodoulides, "Observation of twist-induced geometric phases and inhibition of optical tunneling via Aharonov-Bohm effects," Sci. Adv. 5, 8135 (2019).
61. M. Parto, H. Lopez-Aviles, M. Khajavikhan, R. Amezcua-Correa, D. N. Christodoulides, "Topological Aharonov-Bohm suppression of optical tunneling in twisted nonlinear multicore fibers," Phys. Rev. A. 96, 043816 (2017).
62. I. V. Barashenkov, L. Baker, and N. V. Alexeeva, "PT-symmetry breaking in a necklace of coupled optical waveguides," Phys. Rev. A 87, 033819 (2013).
63. S. Longhi, "PT phase control in circular multi-core fibers," Opt. Lett. 41, 1897 (2016).
64. X. Zhang, V. A. Vysloukh, Y. V. Kartashov, X. Chen, F. Ye, and M. R. Belic, "PT-symmetry in nonlinear twisted multicore fibers," Opt. Lett. 42, 2972 (2017).
65. R. Chen, M. Zou, X. Wang and A. Tennant, "Generation and beam steering of arbitrary-order OAM with time-modulated circular arrays," IEEE Systems Journal 15, 5313 (2021).
66. H. Sakaguchi and B. A. Malomed, "Solitary vortices and gap solitons in rotating optical lattices," Phys. Rev. A 79, 043606 (2009).
67. J. Cuevas, B. A. Malomed, and P. G. Kevrekidis, Two-dimensional discrete solitons in rotating lattices, Phys. Rev. E 76, 046608 (2007).
68. H. Sakaguchi and B. A. Malomed, "Two-dimensional matter-wave solitons in rotating optical lattices," Phys. Rev. A 75, 013609 (2007).
69. S. Jia and J. W. Fleischer, "Nonlinear light propagation in rotating waveguide arrays," Phys. Rev. A 79, 041804 (2009).
70. X. Zhang, F. Ye, Y. V. Kartashov, V. A. Vysloukh, and X. Chen, Localized waves supported by the rotating waveguide array, Opt. Lett. 41, 4106 (2016).
71. C. Milian, Y. V. Kartashov, L. Torner, "Robust ultrashort light bullets in strongly twisted waveguide arrays," Phys. Rev. Lett. 123, 133902 (2019).
72. V. Jarutis, A. Matijošius, P. D. Trapani, and A. Piskarskas, "Spiraling zero-order Bessel beam," Opt. Lett. 34, 2129 (2009).
73. L. Paterson, M. P. MacDonald, J. Arlt, W. Sibbett, P. E. Bryant, and K. Dholakia, "Controlled rotation of optically trapped microscopic particles," Science 292, 912 (2001).
74. Y. V. Kartashov, V. A. Vysloukh, and L. Torner, "Soliton spiraling in optically induced rotating Bessel lattices," Opt. Lett. 30, 637 (2005).
75. P. Zhang, S. Huang, Y. Hu, D. Hernandez, and Z. Chen, "Generation and nonlinear self-trapping of optical propelling beams," Opt. Lett. 35, 3129 (2010).
76. A. Zannotti, F. Diebel, M. Boguslawski, and C. Denz, "Chiral light in helically twisted photonic lattices," Adv. Optical Mater. 5, 1600629 (2017).
77. G. K. L. Wong, M. S. Kang, H. W. Lee, F. Biancalana, C. Conti, T. Weiss, P. St. J. Russell, "Excitation of orbital angular momentum resonances in helically twisted photonic crystal fiber," Science 337, 446 (2012).
78. P. Roth, Y. Chen, M. C. Günendi, R. Beravat, N. N. Edavalath, M. H. Frosz, G. Ahmed, G. K. L. Wong, and P. St. J. Russell, "Strong circular dichroism for the $HE_{11}$ mode in twisted single-ring hollow-core photonic crystal fiber," Optica 5, 1315 (2018).
79. E. V. Barshak, C. N. Alexeyev, B. P. Lapin, M. A. Yavorsky, "Twisted anisotropic fibers for robust orbital-angular-momentum-based information transmission," Phys. Rev. A 91, 033833 (2015).
80. P. St. J. Russell, R. Beravat, and G. K. L. Wong, "Helically twisted photonic crystal fibers," Phil. Trans. R. Soc. A 375, 20150440 (2016).
81. Y. S. Kivshar and G. P. Agrawal, *Optical solitons: From fibers to photonic crystals* (Academic press, Amsterdam, 2003).
82. A. Szameit and S. Nolte, "Discrete optics in femtosecond-laser-written photonic structures," J. Phys. B: Atomic, Molecular and Optical Physics **43**, 163001 (2010).
83. L. J. Maczewsky, M. Heinrich, M. Kremer, S. K. Ivanov, M. Ehrhardt, F. Martinez, Y. V. Kartashov, V. V. Konotop, L. Torner, D. Bauer, A. Szameit, "Nonlinearity-induced photonic topological insulator," Science **370**, 701 (2020).
84. C. Caloz, A. Alù, S. Tretyakov, D. Sounas, K. Achouri, and Z.-L. Deck-Léger, "Electromagnetic nonreciprocity," Phys. Rev. Appl. **10**, 047001 (2018).
85. See Supplementary Material for evolution of linear spectrum of twisted array with other type of discrete rotation symmetry $\mathcal{C}_8$ with increase of the rotation frequency $\omega$, transformation of linear eigenmodes, and stability properties of vortex solitons for different $\omega$ values.
86. Y. V. Kartashov and D. A. Zezyulin, "Stable multiring and rotating solitons in two-dimensional spin-orbit-coupled Bose-Einstein condensates with a radially periodic potential," Phys. Rev. Lett. 122, 123201 (2019).
87. D. Leykam, V. V. Konotop, and A. S. Desyatnikov, "Discrete vortex solitons and parity time symmetry," Opt. Lett. 38, 371 (2013).
88. Y. V. Kartashov, V. V. Konotop, and L. Torner, "Topological states in partially-PT-symmetric azimuthal potentials," Phys. Rev. Lett. 115, 193902 (2015).